\documentstyle[manuscript, natbib]{aastex}

\slugcomment{Not to appear in Nonlearned J., 45.}

\shorttitle{Collapsed Cores in Globular Clusters}
\shortauthors{Djorgovski et al.}

\begin{document}


\title{Signatures of magnetic reconnection at the footpoints of fan shape jets on a light bridge driven by photospheric convective motions  }

\author{Xianyong Bai\altaffilmark{1,2}, Hector Socas-Navarro\altaffilmark{3,4}, Daniel N\'obrega-Siverio\altaffilmark{5,6},Jiangtao Su\altaffilmark{1,2},Yuanyong Deng\altaffilmark{1,2}, Dong Li \altaffilmark{7,8}, Wenda Cao \altaffilmark{9}, Kaifan Ji\altaffilmark{10} }
\altaffiltext{1}{National Astronomical Observatories, Chinese Academy of Sciences, 20 Datun Road, Beijing 100101, China; xybai@bao.ac.cn}
\altaffiltext{2}{Key Laboratory of Solar Activity, National Astronomical Observatories, Chinese Academy of Sciences, 20 Datun Road, Beijing 100012, China}
\altaffiltext{3}{Instituto de Astrof\'isica de Canarias, Via L\'actea, s/n, E-38205 La Laguna (Tenerife), Spain}
\altaffiltext{4}{Department of Astrophysics, Universidad de La Laguna, E-38200 La Laguna (Tenerife), Spain}
\altaffiltext{5}{Institute of Theoretical Astrophysics, University of Oslo, P.O. Box 1029 Blindern, NO-0315 Oslo, Norway}
\altaffiltext{6}{Rosseland Centre for Solar Physics, University of Oslo, P.O. Box 1029 Blindern, NO-0315 Oslo, Norway}
\altaffiltext{7}{Purple Mountain Observatory,Chinese Academy of Sciences, No.8 Yuanhua Road, Qixia District, Nanjing 210034, China }
\altaffiltext{8}{Key Laboratory of Dark Matter and Space Astronomy, Purple Mountain Observatory, CAS,
Nanjing 210034, China }
\altaffiltext{9}{Big Bear Solar Observatory, New Jersey Institute of Technology, Big Bear City, CA 92314-9672, U.S.A.}
\altaffiltext{10}{Yunnan Observatories, Chinese Academy of Sciences, P.0.Box110, Kunming 650011, Yunnan, China}

\begin{abstract}
Dynamical jets are generally found on Light bridges (LBs), which are key to studying sunspots decays. So far, their formation mechanism is not fully understood. In this paper, we used state-of-the-art observations from the Goode Solar Telescope, the Interface Region Imaging Spectrograph, the Spectro-Polarimeter on board Hinode and the Atmospheric Imaging Assembly (AIA) on board the Solar Dynamics Observatory to analyze the fan shape jets on LBs in detail. Continuous upward motion of the jets in ascending phase is found from the H$\alpha$ velocity, which lasts for 12 minutes and is associated with the H$\alpha$ line wing enhancements. Two mini jets appear upon the bright fronts of the fan shape jets visible in the AIA 171 {\AA} and 193 {\AA} channels, with a time interval as short as 1 minute. Two kinds of small scale convective motions are identified in the photospheric images, along with the H$\alpha$ line wing enhancements. One seems to be associated with the formation of a new convection cell and the other manifests as the motion of a dark lane passing through the convection cell. The finding of three lobes Stokes V profiles and their inversion with NICOLE code indicates that there is magnetic field lines with opposite polarities in LBs. From the H$\alpha$ -0.8 {\AA} images, we found ribbon like brightenings propagating along the LBs, possibly indicating slipping reconnection. Our observation supports that the fan shape jets under study are caused by the magnetic reconnection and photospheric convective motions play an important role in triggering the magnetic reconnection.
\end{abstract}


\section{INTRODUCTION}

Light bridges (LBs), which commonly occur during the sunspot decay phase, are bright and elongated structures that separate umbra cores or are embedded in the umbra. Studying magnetic and dynamical properties of LBs is very helpful for the understanding of sunspots decay. According to the brightness and size of LBs, they can be categorized into three types: faint, strong and granular LBs. Faint LBs have an elongated structure composed of grains of similar size and shape to umbral dots \citep{Sobotka1993,Sobotka2009}. Strong LBs, whose typical brightness is comparable to the penumbra, separate the umbra into two regions with the same polarity \citep{Rimmele2008,Rezaei2012}. Granular LBs show convection cells similar to the granules in the quiet sun \citep{Vazquez1973,Sobotka1994,Leka1997,voort2010}.

From observations, it has been found that all LBs have a relative weaker magnetic field strength and an increased magnetic inclination compared with the surrounding umbra \citep{Leka1997, Jur2006,Katsukawa2007}, forming a cusp-like magnetic field configuration at higher layers. Due to the magnetic shear between the horizontal fields of LBs and the vertical fields of umbra, a sharp and strong current layer forms at the edges of the LBs, which is a favorable site for magnetic reconnection \citep{Shimizu2009, Toriumi2015, Liu2015}. From the line-of-sight (LOS) velocity images, central hot upflows are found in LBs surrounded by cooler fast downflows, sometimes at supersonic speeds \citep{Rimmele1997, Louis2009, lagg2014, Felipe2016, Schlichenmaier2016}, supporting their convective origin. In addition, \citet{Toriumi2015} discovered some small-scale, short-lived convection cells superimposed on the large-scale, long-term velocity structures in the LBs in the formation phase of an active region (AR). Some faint divergent magnetic patterns are also found, which suggests continuous supply of weak magnetic flux from the solar interior, transported by large-scale upflows of the LBs \citep{Louis2015}. \citet{Felipe2016} showed that the value of plasma $\beta$ in LBs is in the range of 1 to 200, indicating that the LBs are dominated by plasma motions.

In the upper atmosphere upon LBs, many dynamical activities are found, which comprise multi-temperature and multi-disciplinary structures \citep{Rezaei2018}. For example, brightenings are very common on broadband images in the wavelength of the Ca {\sc ii} H, AIA 1600 {\AA}, 1700 {\AA}, IRIS 1300 {\AA}, 1400 {\AA}, and Mg {\sc ii} 2796  {\AA} \citep{Shimizu2009, Berger2003, Rezaei2018}. Also there are lots of long lasting and recurring jets (or surges) which are seen with the cool chromospheric lines such as the H$\alpha$ and Ca {\sc ii} lines \citep{Roy1973, Asai2001, Bharti2007, Liu2012, Louis2014, Robustini2016,Song2017}. \citet{Bharti2015} and \citet{Yang2015} reported that most of the jets seen in the chromospheric images show decreasing brightness with height while they show a bright front in the IRIS 1330 {\AA} slit-jaw images (SJI). Some of the bright fronts also exist in the AIA 171 and 131 {\AA}  channels, suggesting that the jets can be heated up to transition region (TR) and low coronal temperatures \citep{Yang2016,Yuan2016,Houjy2016}. Moreover, the jets' bright fronts seen in IRIS 1400 {\AA} on LBs are disturbed by their surrounding activities, e.g., adjacent brightenings and solar flares \citep{Hou2016,Yang2017,Felipe2017}.

Two mechanisms for the formation of the jets on LBs are generally supported by observations. One is driven by magnetic reconnection, evidenced by the high speeds ($100$ km$\ s^{-1}$) of the jets and current layers at the edge of the LBs \citep{Toriumis2015,Toriumi2015}. The other is caused by shocked p-mode waves leaked from the underlying photosphere, with lower speeds and nearly stationary oscillating period with several minutes \citep{Yang2015,Hou2017, Zhang2017}. \citet{Tian2018} shows that the jets (or surges) on LBs has two components according to their occurrence frequency and maximum length. The ever-present short surges seems to be related to the upward leakage of magnetoacoustic waves from the photosphere, while the occasionally occurring long and fast surges are obviously caused by the intermittent reconnection. All the above mentioned observations revealed the high dynamical nature of the LBs and their complex interaction with the adjacent environments.

From the theoretical perspective, radiative magnetohydrodynamics (MHD) simulations show their advantage in understanding the detailed physical processes behind the various dynamical events on LBs. \citet{Cheung2010} highlighted the common magneto-convective origin of LBs, umbral dots, and penumbral filaments during the formation of an AR. Based on the work by \citet{Cheung2010}, \citet{Toriumis2015}  found that LBs take root deep down in the convection zone and have a large-scale convective upflow, which is consistent with the observations. The upflow carries part of the weak horizontal fields in deeper layer to the solar surface. Due to the radiative cooling, the ascending plasma loses buoyancy and turns back into the convection zone at the edge of LBs lanes, showing downflows. At the LBs' boundaries, a strong electric current layer is formed, providing the energy for various dynamical events. Even though the numerical experiments have been helpful providing theoretical support for the observations and understand some key features of LBs, there is no radiative magnetohydrodynamics simulations of the jets in the upper layers of LBs.

Some questions are still unresolved about the dynamical jets on LBs, in particular, for the occasionally occurring long and fast jets. If they are caused by magnetic reconnection, then what is the trigger mechanism for the reconnection? Could we obtain more information during the magnetic reconnection? What is the relationship between the dynamical jets and their underlying photospheric evolution of mass flows and magnetic fields? Thanks to high resolution coordinated observations from ground and space, we are in an advantageous position to answer those open questions.

In this paper, we focus on the fan shape jets on LBs, which is generally found in the H$\alpha$ off-band images. Similar jets are reported by \citet{Asai2001} and \citet{Robustini2016}. The aims are to reveal the fine structures of the photospheric convection on LBs and to find the relationship among the small-scale photospheric flow, brightenings and the jets like activities on their upper layers, which is helpful for us to understand the coupling between mass and energy on LBs. To that end, we have used joint observations from the the Goode Solar Telescope (GST, previously called New Solar Telescope) \citep{Cao2010}, the Interface Region Imaging Spectrograph (IRIS) \citep{De2014}, the Spectro-Polarimeter (SP) instrument on board Hinode \citep{Lites2013} and the Atmospheric Imaging Assembly (AIA) instrument on board the Solar Dynamics Observatory (SDO) \citep{Lemen2012}. This way, we have been able to cover the different layers of the solar atmosphere, namely, the photosphere, chromosphere, TR and corona.

The paper is organized as follows. In Section \ref{sec2}, the data from both ground and space based solar telescopes and the way to reduce them are presented. Section \ref{sec3} shows the results followed by a summary in the last section (Section 4).

\section{OBSERVATIONS AND DATA REDUCTION}
\label{sec2}

We have used coordinated observations of GST, IRIS and HINODE/SP obtained on 2014 August 1 from 17:15 to 17:55 UT. The pointer was centered at heliocentric coordinates (x,y)=(-192", -210"), targeting at the decaying sunspot of AR 12127. The GST data contain simultaneous observations of the photosphere, using the titanium oxide (TiO) line taken with the Broadband Filter Imager, and the chromosphere, using the H$\alpha$ 6563 {\AA} line obtained with the Visible Imaging Spectrometer (VIS). The passband of TiO filter is 10 {\AA}, centered at 7057{\AA} while its temporal resolution is about 15 s with a pixel scale of  $0.0343\arcsec$. Concerning VIS, a combination of 5 {\AA} interference filter and a Fabry-P\'erot etalon is used to get a bandpass of 0.07 {\AA} at the H$\alpha$ line. Its field of view (FOV) is about 70$^{''}$ with a pixel scale of 0.0323$^{''}$. To obtain more spectral information, we scan the H$\alpha$ line at 11 positions with a 0.2 {\AA} step following this sequence: $\pm 1.0, \pm 0.8, \pm 0.6, \pm 0.4, \pm 0.2, 0.0\ ${\AA}. The time used in each scan is 23 s. Both high-order adaptive optics \citep{Cao2010} and speckle reconstruction post-processing method \citep[KISIP speckle reconstruction code;][]{Woger2007} are employed to achieve diffraction-limited resolution images.

IRIS data are taken in a very large coarse 64-step raster mode, with a step size of 0.35$^{''}$, step cadence of 31.8 s and a raster cadence of 2036 s. The exposure time at each step is 29 s. Several strong emission lines are recorded, such as the Si {\sc iv} 1402.77 {\AA}, Si {\sc iv} 1393.7 {\AA}, C {\sc ii} 1334.53 {\AA} and Mg {\sc ii} k 2796.35 {\AA}. The cadences of the SJI in the passbands of 1400 {\AA}, 2796 {\AA}, and 2832 {\AA} passband are all 112 s. The spatial resolution is about $0.33^{''}$ and the FOV is 22.4$^{''}$$\times$175$^{''}$ and 175$^{''}$$\times$175$^{''}$ corresponding to the 64-step raster mode and the SJI. All the spectra and images used in the analysis are level 2 data, including dark current subtraction, flat-field and geometrical corrections \citep{De2014, Tian2014}.

Hinode/SP data are acquired in a fast mode with a FOV of 81$^{''}$$\times$81$^{''}$, a step size of 0.297$^{''}$ and a raster cadence of 1080 s. Additionally, in order to analyze the corona response to our event, we have used SDO data, including the EUV and UV passbands from the AIA \citep{Lemen2012} with the cadences of 12 s and 24 s, repectively.

Spatial register between the GST and SDO is first done by co-aligning the AIA 1600 {\AA} images with the GST/TiO data, and then co-align the TiO data with the H$\alpha$ images at different wavelength positions. We use the AIA 1600 {\AA}, IRIS 2832 {\AA} and SP continuum images to align the SDO, IRIS and SP data. The uncertainty of the coalignment is about one AIA pixel (0.6$^{''}$). Figure \ref{fig1} contains the context image of the LBs and fan shape jets on the sunspot of AR 12127 that we analyze in the subsequent section. The figure shows data from GST, IRIS and SDO/AIA after the alignment process.

\section{ANALYSIS AND RESULTS}
\label{sec3}
\subsection{The realitionship between H$\alpha$ brightenings at the footpoints of fan shape jets and their bright fronts shown in the AIA EUV channels }

\begin{figure}
\epsscale{1.0} \plotone{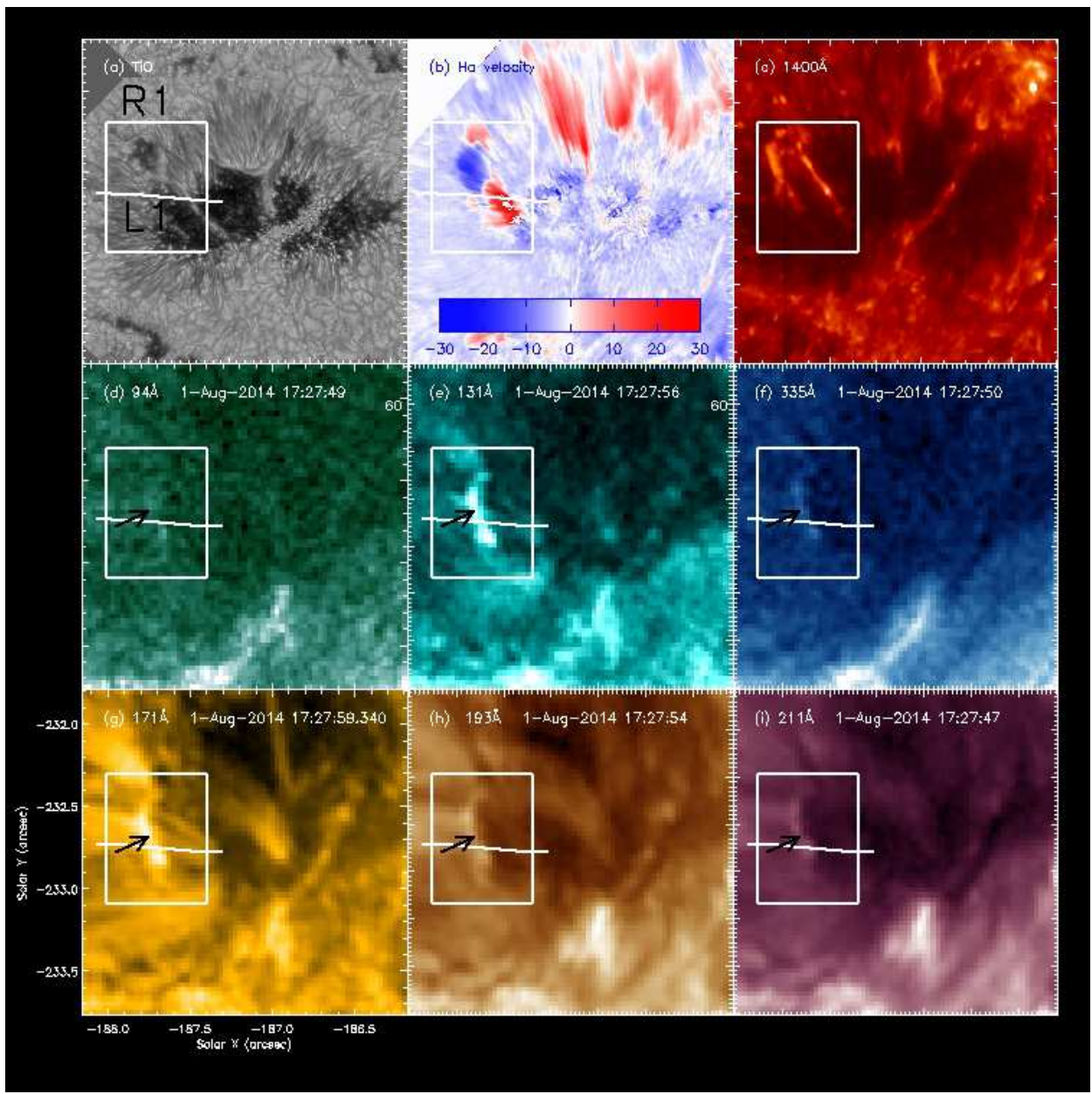} \caption{ Context image of the LBs and fan shaped jets on the sunspot of the AR 12127 obtained on 2014 August 1, around 17:27 UT. The different panels show (a) the photosphere through TiO images from GST; (b) chromospheric via Ha LOS velocity obtained with GST/VIS; (c) The TR through SJI 1400 images from IRIS; and (d-i) the corona response observed with the different filters of SDO/AIA. R1 illustrates the LB region, to be analysed in Fig. \ref{fig3}. L1 represents a slice to be used in Fig. \ref{fig2}. Black arrow in panel (d)-(i) marks bright fronts of the fan shape jets seen in 171 {\AA}.  An associated animation (movie1.avi) is available online. }
\label{fig1}
\end{figure}

The fan shape jets on the LB can be found in the region R1 in figure \ref{fig1}b and \ref{fig1}c, corresponding to the velocity derived from the H$\alpha$ line using 11 points and the IRIS SJI in the 1400 {\AA} passband, respectively. The blue and red colors in the LOS velocity images indicate the plasma showing upward and downward motions on the fan shape jets in relative cool temperature (less than 10000 K). The associated online animation clearly reveals this evolution of upflow and downflows. Signatures of the fan shape jets can also be identified in the IRIS 1400 {\AA} passband and the six EUV channels of AIA, i.e. 94 {\AA}, 131 {\AA}, 171 {\AA}, 193 {\AA}, 211 {\AA}, and 335 {\AA}, indicating that they have hot materials, covering from 10$^5$ K to 10$^6$ K. The cool and hot components seen from the simultaneous multi-wavelength observation reveal multi-thermal nature of the fan shape jets. Moreover, the jets' fronts show brightenings (region marked by the arrow in Fig. \ref{fig1}d-\ref{fig1}i)) in the above mentioned passbands while the footpoint of the jets on LB has some transient brightenings, especially in the IRIS 1400 {\AA} image.

\begin{figure}
\epsscale{1.0} \plotone{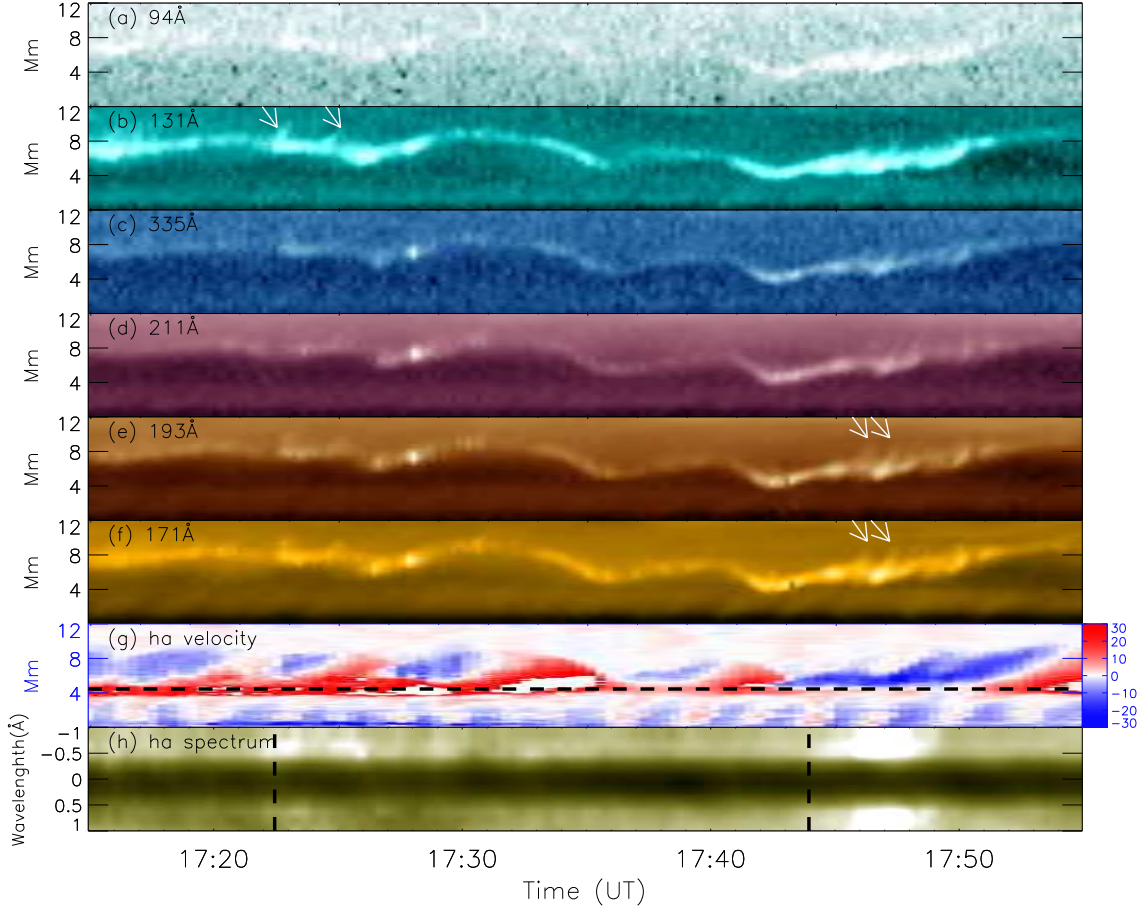} \caption{ Space-time map along the slice L1 shown in Fig. \ref{fig1}. Panels $(a)$ $-$ $(f)$ show the maps for the different filters of AIA, namely, 94, 131, 335, 211, 193, 171 {\AA}. Panel (g) contains the LOS velocity of the H$\alpha$ line. Panel $(h)$ shows the evolution of the H$\alpha$ spectrum at the location marked by the horizontal dashed line of panel $(g)$. Two white arrows near 17:47 UT mark the two mini jets occurring at a time interval of about 60 s upon the bright jets' front. The two vertical dashed lines in panel (h) mark the two times associated with the H$\alpha$ line wing enhancements and the beginning of blue shifts in H$\alpha$ LOS velocity. }
\label{fig2}
\end{figure}

Along the trajectory of the jets, a slice, labelled as L1, is selected to study the dynamical response of the different solar atmosphere layers to the jets. The corresponding space-time diagram is presented in figure \ref{fig2}. The first we notice is that the jets follow a parabolic trajectory. This is evident following the shape of the brightenings (or bright fronts) observed in the six EUV channels of AIA (Figure \ref{fig2}, panels a to f), e.g., from 17:27 to 17:36 UT. Evolution of the LOS velocity in the H$\alpha$ line in Fig. \ref{fig2}g reveals that the jets have blueshift (upward motion) at the beginning, and then they gradually become redshift (downward motion) later. Combining with the parabolic trajectory in the plane of the sky (POS), one can imagine that the jets leave the solar surface with a high initial speed, decrease their velocity due to the gravity, reach their maximum height and finally fall back to the solar surface again. The maximum speeds in the LOS and POS direction at the beginning of the jets eruption are about 25 km$\ s^{-1}$ and 15 km$\ s^{-1}$ near 17:27 UT. The corresponding values near 17:43 UT are about 30 km$\ s^{-1}$ each. The two times are selected because line wing enhancements in the H$\alpha$ line are found in the meantime. The total speeds (square root of the summation of LOS velocity's square and POS velocity's square) of the launched jets at the two times are about 29 and 42 km$\ s^{-1}$, which are much larger than the typical local sound speed in the chromopshere (less than 10 km$\ s^{-1}$). The large initial velocity values are likely to promote shock waves, so the bright fronts visible in IRIS and SDO can be interpreted as the effect of shock compression of TR's and coronal material as the cool jet slams into it \citep{Robustini2016}. The fast nature of shocks can also produce departures from statistical equilibrium in the different populations of the emitting ions. Therefore, enhanced brightenings in the bright front of the jets could also be related to non-equilibrium effects as it occurs in surges due to the quick action of entropy sources \citep{NS2018}.

Figure \ref{fig2}h shows temporal sequences of the H$\alpha$ spectrum of VIS at the location marked by the dashed line in figure \ref{fig2}g. Both wings of the H$\alpha$ spectrum is extremely enhanced near 17:24 and 17:47 UT while the line core keeps unperturbed, similar to the profiles of Ellerman bombs (EBs) \citep{Ellerman1917,Robustini2016,Song2017,Chen2017,Hong2017,Hong2017ApJ}. Generally, the appearance of EBs spectra indicate that magnetic reconnection is happening in the upper photosphere or the lower chromosphere, namely near the temperature minimum region. Comparing figure \ref{fig2}h with \ref{fig2}g, one can find that the line wing's enhancements in the H$\alpha$ spectrum are associated with blueshift as shown in the LOS velocity, especially near 17:47 UT. At the same time, all the EUV channels also show brightenings. The velocities found in the jets, together with the enhancements of the wings in the H$\alpha$ spectrum, seem to indicate that the fan shape jets studied in this paper are caused by magnetic reconnection.

It is worth mentioning that the LOS velocity in the H$\alpha$ line (Figure \ref{fig2}, panel g) shows continuous blueshift from 17:42 to 17:54 UT, lasting for 12 minutes. Meanwhile, wave or oscillation like patterns occur four times at the base of the slice L1 covering the LB and adjacent umbra, evidenced by the alternating appearance of the blueshifts and redshifts. The long time interval of the blue shift in the jets possibly indicates that they are not driven by the oscillations. The brightenings in the bright fronts of jets in the AIA 171 {\AA}, 193 {\AA} and 211 {\AA} channels have some fine structures, manifested as two mini jet-like (marked by the two arrows in Figure \ref{fig2}f) brightenings on the background of jets' fronts near 17:47 UT, together with the brightening wings in H$\alpha$ line. The interval of the two jets is about 1 minute. Moreover, similar two mini jet-like structures with the interval about 2 minutes can also be found near 17:23 UT (marked by the two arrows in Figure \ref{fig2}b), which are most clearly seen in the 131 {\AA} channel, but are present in the others. The mini jets in such a short time interval are not possibly caused by the leakage of waves, supporting its correlation with magnetic reconnection. These mini and short lived jets possibly reveal some fine structures during the magnetic reconnection process.

The line wing enhancement in the H$\alpha$ line occurs not only at the point shown in figure \ref{fig2}h, but also at other points on the footpoints of the jets located on LB. Figure \ref{fig3} presents the brightenings on LB observed at the -0.8 {\AA} from  H$\alpha$ line center for different stages of the evolution. From figure \ref{fig3}a to \ref{fig3}c, they show that the jets move to higher layers evidenced by the increasing length of the jets along their propagation direction. Note that the footpoints of the jets get bright which look like a ribbon at 17:20:38 UT. From Fig. \ref{fig3}d to \ref{fig3}f, one can find that the ribbon like brightening propagates along the LB at a speed of about 9.46 km$\ s^{-1}$, indicated by the green solid line. Meanwhile, the jets on the LB move in the same direction with the brightenings (marked by the arrows), which is almost perpendicular to the jets' propagation direction. According to Fig. \ref{fig3}g to \ref{fig3}i, the jets occur firstly at both ends of the LB with remarkable brightenings at their footpoints. Then, the brightenings at the both ends move in opposite direction with the speed of about 30 $km\ s^{-1}$. At 17:48:53 UT, one can also see a long ribbon appearing along the entire LB (see the online movie 2).  The jets launched along the entire LB soon afterwards. The ribbon motions seem to be the signatures of slipping reconnection, which is generally found in the flare ribbons \citep{Aulanier2006, Li2014, Li2016, Jing2017}. The result reveals the three-dimensional nature of the fan shape jets. The magnetic reconnection occurs successively at different locations on the LB so that we can see the brightening motions. The width of the ribbon is about 0.3~$^{''}$, much smaller than the spatial resolution of AIA.

\subsection{TR's spectrum at the footpoint and bright front of the fan shape jets}

In this coordinated campaign, IRIS merely has one scan during time interval of observations with 64 steps. The upper panels of figure \ref{fig4} show the slit scanning image near the line centers of Si {\sc iv} 1402.77 {\AA} and Mg {\sc ii} K2v line at 2795.94 {\AA}. Similar to the bright features of SDO/AIA mentioned above, both the footpoints and the front of the jets are also bright in the Si {\sc iv} 1402.77 {\AA} and Mg II K2v. Note that although Mg {\sc ii} K2v and H$\alpha$ are both chromospheric lines, the latter does not show any bright fronts, possible due to the increase of source function at the selected wavelength position \citep{Leenaarts2013}. Those brightenings suggest that there is heating of chromospheric and TR region plasma on the jets.

In order to analyze the IRIS spectral information, we choose three points on the bright fronts of the fan shape jets (marked by P1, P2 and P3) and two points on its footpoints (P4 and P5). Their corresponding spectrum in the C {\sc ii} 1335, Si {\sc iv} 1394 and 1403 {\AA} is shown in the panels of the lower two rows of figure \ref{fig4}. The black line is the reference spectrum from the average of the bottom rows of the slit scanning image with a region of 12$^{''}$ $\times$ 2$^{''}$. The three points on the jets' bright fronts show three different types of spectrum and all of them have line broadening relative to the reference spectral profiles, as seen from the Si {\sc iv} 1403 and 1394 {\AA} lines. The green line, located at the brightest region (P2), has the highest emission. The blue line for P3, whose emission is at the intermediate level, shows strong blue asymmetry in all the lines of Si {\sc iv} 1397, 1403 and C {\sc ii} 1335 {\AA} characterized by multi-components profiles. The red line located at P1 has the weakest emission among the three points and manifests as two gaussian emission structures, which is an indication of plasma motions with two components. This idea could be supported by the results of \citet{NS2017, NS2018}. In those papers, the authors show that for surges, the alignment of the LOS with the orientation of the ejections is key to understanding the remarkable TR brightenings as well as the multi-component profiles. This result for surges could be also applied for our jets, so the multi-component profiles found may be due to the integration of the multiple-crossing emitting layers of the folded TR on the fan shape jets with different LOS velocities.

Concerning the footpoints of the fan shape jets (P4 and P5 in Figure \ref{fig4}), they show line broadening and intensity enhancement. We also find absorption structures of Ne {\sc ii} 1393.33 and 1335.2 {\AA} superimposed on the lines of Si {\sc iv} 1393.76 {\AA} and C {\sc ii} 1335 {\AA} especially for P5 (green line) with a higher emission. The appearance of the absorption features in enhanced TR spectral profiles is an indication of a hot reconnection region below the dense and cold plasma \citep{Peter2014,Tian2018}. We also checked the H$\alpha$ images at -0.8 {\AA} from the line center and found that there is intermittent intensity enhancements at the same time. These observational characteristics also support that the fan shape jets are caused by the magnetic reconnection. Note however that the absorption features observed here are not akin to the one discussed by \citet{Tian2018}. In their paper, significant line broadening of the Si {\sc iv}, C {\sc ii} and Mg {\sc ii} $h\&k$ lines are found and the Mg $h\&k$ lines show strong absorption due to large opacity. Almost all of the lines have absorption line structures. In our data, no significant absorption structures are found in the Si {\sc iv} 1403 {\AA} line, possibly due to the emission at P4 and P5 being relative weaker.

\subsection{Photospheric large and small scale convection flows on LBs driving the magnetic reconnection}

Concerning Hinode/SP data, two scans were performed from 17:15 to 17:55 UT. Figure \ref{fig5}a shows the longitudinal magnetic field in level 2 for Region R1 as shown in Fig. 1a. It can be found that its value is much lower than that in the pore or umbra. The arrows superimposed in the panel indicate the strength and direction of the transverse magnetic field. The transverse field changes its directions on LB, indicating the existence of magnetic shear.  The current sheet is formed at the magnetic shear region, storing enough non-potential energy for the fan shape jets \citep{Toriumis2015}. We also obtained the inclination angles on LB and showed them in Fig. \ref{fig5}b. The inclination angles, with the value close to 90$\degr$, are much larger than those in the the nearby pores and umbra, revealing an almost horizontal magnetic structure.

We derived the horizontal velocity on the LB with the high resolution TiO images from GST following the method explained by \citet{Liu2018}. The result is arranged in figure \ref{fig5}c and \ref{fig5}d, where the corresponding background images are the SP LOS velocity and the TiO image, respectively. The presented horizontal velocity corresponds to the time of 17:27 and 17:44 UT. When plotting the horizontal velocity, an running average of the ones calculated from \citet{Liu2018}'s method with a time interval of 100 s is adopted so as to reduce the noise. Two large scale flow structures occur during the 40 minutes' observation (see the online movie 3). They moves in the opposite direction from almost the center of the LB, with its upper left to one end of the LB and that in lower right to the other end of the LB. The maximum speed of the horizontal flow is about 1.8 km$\ s^{-1}$. From the background image in Fig. \ref{fig5}c, one can also find that the upper left part of the LB has a upflow being blueshift while the lower right part has a downflow. In the upper left part (region R2), the plasma on the LB shows upward motion and changes to downward motions at the boundary between the LBs and the nearby pore, indicating the convective motions of the LBs.

The small scale photospheric convection on the LB can be found from the evolution of the TiO images in region R2, covering the upper left parts of the LB. The result is shown in Fig. \ref{fig6}. From 17:20 to 17:25 UT, one can find some fragmental structures near the location of P6. At the time of 17:27, they develop into a new convection cell which expands after comparing with the subsequent images. In addition, the direction of horizontal velocity in the region as shown in Fig. \ref{fig5}c shows that the flow moves from the cell center to its boundary. Combing the upflows in the LOS velocity, it indicates the emergence of a new convection cell. Intensity enhancements at 17:27 UT near P6 on TiO image are obvious. The red contour marks the brightenings seen from the simultaneous H$\alpha$ -0.8 {\AA} passband. It indicates that there is response in the photosphere during the EBs, probably related to the magnetic reconnection heating. From 17:40 to 17:51 UT, the motion of a dark lane (marked by the arrow) is notable. It first occurs near the boundary of the penumbra and the LB, then moves to the upper left direction, and finally reaches at the upper boundary of the LB. At the time of 17:49 and 17:51 UT, one can also find the intensity enhancements near the dark lane on the TiO images, which is accompanied with the brightenings in the H$\alpha$ -0.8 {\AA} passband. The scenario shows that the small scale photospheric convection plays important roles in triggering the EBs and subsequent fan shape jets. Recently, \citet{Pasechnik2018} also reported an increase in the core intensity of the studied photospheric lines, correlated spatially with the enhanced emission in the H$\alpha$ line wing.

\subsection{Abnormal Stokes V profiles on LBs associated with the H$\alpha$ line wing brightenings }

From the SP longitudinal magnetograms in level 2, we did not found small scale magnetic structures with opposite polarity on the LB. However, abnormal Stokes V profiles do exist in the level 1 data. The location of the abnormal Stokes V profiles is marked by P6 and P7 in Fig. \ref{fig6}. The corresponding Stoke I, Q, U and V profiles can be found in Fig. \ref{fig7}, acquired at 17:27 and 17:44 UT, respectively. Note that there is an enhanced emission at the H$\alpha$ -0.8 {\AA} passband near the two times, possibly indicating the line wing brightening has a close connection with these abnormal Stokes V profiles. The asymmetry in the red wings of Stokes I profiles is obvious in Fig. \ref{fig7}. The abnormal Stokes V profiles have three lobes, which is an indicator of the complexity of the LB magnetic field configuration. Taking a look at the TiO images, we can find that there are fine structures in both P6 and P7 regions, pointing out again to the highly complex LB environment. Similar abnormal Stokes V profiles has also been found in the penumbra \citep{Rubio2004,Ruiz2013,lagg2014,Franz2016}, which is thought to be the indication of returning magnetic flux. \citet{Louis2014} and \citet{Felipe2017} also reported the Stokes V profiles with multiple lobes on the LBs. The three lobes in Stokes V profiles are generally interpreted as the signature of opposite polarity in the LOS direction.

In order to fit the observed Stokes I, Q, U and V profiles, we employ NICOLE code \citep{Hector2015}, which is very powerful in dealing with these abnormal profiles thanks to its flexibility to accommodate complex atmospheric models. The level 2 data of Hinode/SP is generated with the analytical solution of the polarized radiative transfer equation under the assumption of Milne¨CEddington atmosphere model, which is unable to generate the asymmetric profiles. The solid lines in Fig. \ref{fig7} are the best fitting results. In total, 8 nodes in the temperature, 4 nodes in the LOS bz and 4 nodes in the LOS velocity are used in the inversion. The LOS bx and by have 1 node. NICOLE can reproduce most of the observing profiles, especially the three lobes in Stokes V profiles. We present the output atmospheric parameters from NICOLE in Fig. \ref{fig8}.

The error bars in the figure have been computed with a MonteCarlo approach, by performing many inversions with different initializations. The errors reflect the statistical spread of the solutions that converged to a fit of similar quality to the best inversion. This approach is optimal to study the uncertainties due to
line sensitivity or possible ambiguities in the solution (possible uniqueness issues), it is constrained to the specific model scenario chosen in this work. For example, it does not rule out other possible solutions obtained with a larger number of nodes or a more complex scenario.

The fitted temperature at P6 covering the enhanced emission at the TiO images also shows an temperature enhancement near the height of 250 km reconfirming the above conclusion that the photosphere is disturbed during the EBs. The temperature at P7 also has an temperature enhancement, but the value is about 500 K lower than that at P6. From the lower panel in Fig. \ref{fig6}, P7 contains the dark lane and its nearby convective cells with higher emission. The fitted temperature at P7 possibly indicates the hot (the temperature enhancements) and cool (dark lane) components. The bx and by are 0 and 600 G at P6 while their values are 320 and -800G at P7, respectively. The key point in the generating the three lobes in the Stokes V profiles is to add stratification with height for the bz and LOS velocity. The bz in the low height has a negative value while it changes to positive values in the upper height. The scenario reveals that there are mixed polarities at P6 and P7. Regarding the LOS velocity, it manifests as redshift in low height while near zero in the upper height. The red shift (downflow) corresponds to negative polarity, contrary to the adjacent magnetic fields. The scenario is possibly consistent with the result from \citet{lagg2014}, where it shows that the downflowing material at the boundary of the LB is able to drag down magnetic field lines, creating a region where additional heating might occurs. \citet{Felipe2016} also claimed that the LBs are plasma-dominated region with the plasma $\beta$ larger than 1 and the convective flow observed on the LBs is able to bend the magnetic field lines to produce field reversals.

\section{SUMMARY AND DISCUSSION}

The joint observations of the fan shape jets on a LB in the decay phase of AR 12127 give us an excellent opportunity to investigate their dynamic behavior in detail. Part of our results are consistent with previous report.  For example, fan shape jets have the bright fronts and intermittent brightenings at the footpoints. The emission of the TR lines has line broadening and intensity enhancements. The appearance of the Ne {\sc ii} absorption feature superimposed on the Si {\sc iv} 1393.76 {\AA} and C {\sc ii} 1335 {\AA} lines at the footpoints of the jets. Furthermore, the present work shows some new observational characteristics, which are summarized as follows.

\begin{enumerate}
  \item The brightest intensity on the front of jets seen in the AIA EUV channel is accompanied by the initiation of upward motion from the H$\alpha$ LOS velocity. Around the time, the H$\alpha$ line wings show intensity enhancements while that at the line center are not obvious, being an indication of EBs. One of jets shows continuous upward motion lasting for 12 minutes. On top of the bright jets' fronts, two small scale jet-like structures also exist with very short lifetime and a time interval as short as 60 s. These results support that the fan shape jets are caused by magnetic reconnection, especially for the occurrence of the small scale jets.

  \item Ribbon-like brightenings are found at the footpoints of LB in the H$\alpha$ -0.8 {\AA} passband images. They also move along the LB, probably showing the scenario of slipping reconnection. It reflects three-dimensional nature of the fan shape jets. In addition, the two-components profiles in the Si {\sc iv} 1403 and 1393 {\AA} lines at the bright jets' fronts confirmed the multiple-crossing effect from the simulation of \citet{NS2017,NS2018}.

  \item There are continuous large scale motions on the LB. The LOS velocities are opposite along the LB, similar to the Evershed flow. Parts of the horizontal flows are to one end of the LB (its nearby umbra) while the other flows are moving to the other end (the pore structures) in an opposite direction. Beside the large scale motions, small scale motions also exist, as presented in Fig. \ref{fig6}.  Evolution of the TiO images shows two kinds of convective motions. One seems to be the emergence of a new convective cell evidenced by the low resolution blue shift and the horizontal divergency flows. The other manifests as a dark lane passing through the convection cells of the LB. The dark lane is formed at the boundary of the LB and its nearby penumbra. We are not sure whether it is a moving magnetic structure due to the lack of magnetograms with high spatial and temporal resolution. Line wing brightenings are found in the H$\alpha$ line associated with these photospheric motions, indicating that the convection motions at the photosphere play important roles in generating the fan shape jets on LB.

  \item The level 1 data with a fast mode from Hinode/SP shows abnormal line profiles on LB, i.e., three lobes in the Stokes V profiles and red asymmetry in the Stokes I profiles. Employing NICOLE code, we are able to reproduce these profiles, which are likely produced by mixed polarities and velocity gradient along the LOS direction. The location of the abnormal Stokes profiles is near the boundary of the LB, where the configuration of magnetic fields is very complicated. The convective motions in the convection cells are possible to carry out some small scale magnetic fields to the edge of the LB and interact with the preexisting sunspot magnetic fields, causing the magnetic reconnection. However, due to the limited resolution of the Hinode/SP, it is difficult to find direct evidences of the opposing polarities in the convection cells of LB. With the high spatial and temporal resolution vector magnetograms of GST, it is possible to find more evidences of the magnetic fields with mixed polarities.
\end{enumerate}

Combining with the small scale photospheric motion, the abnormal Stokes profiles, the appearance of mini jets with short interval upon the bright jets' fronts in the AIA channels, the H$\alpha$ line wing enhancements, the motions of ribbon-like brightenings at the H$\alpha$ -0.8 {\AA} passband and the Ne {\sc ii} absorption feature in the TR lines, we claim that the fan shape jets on the LB are due to the magnetic reconnection at the upper photosphere or lower chromosphere.

\begin{figure}
\epsscale{1.0} \plotone{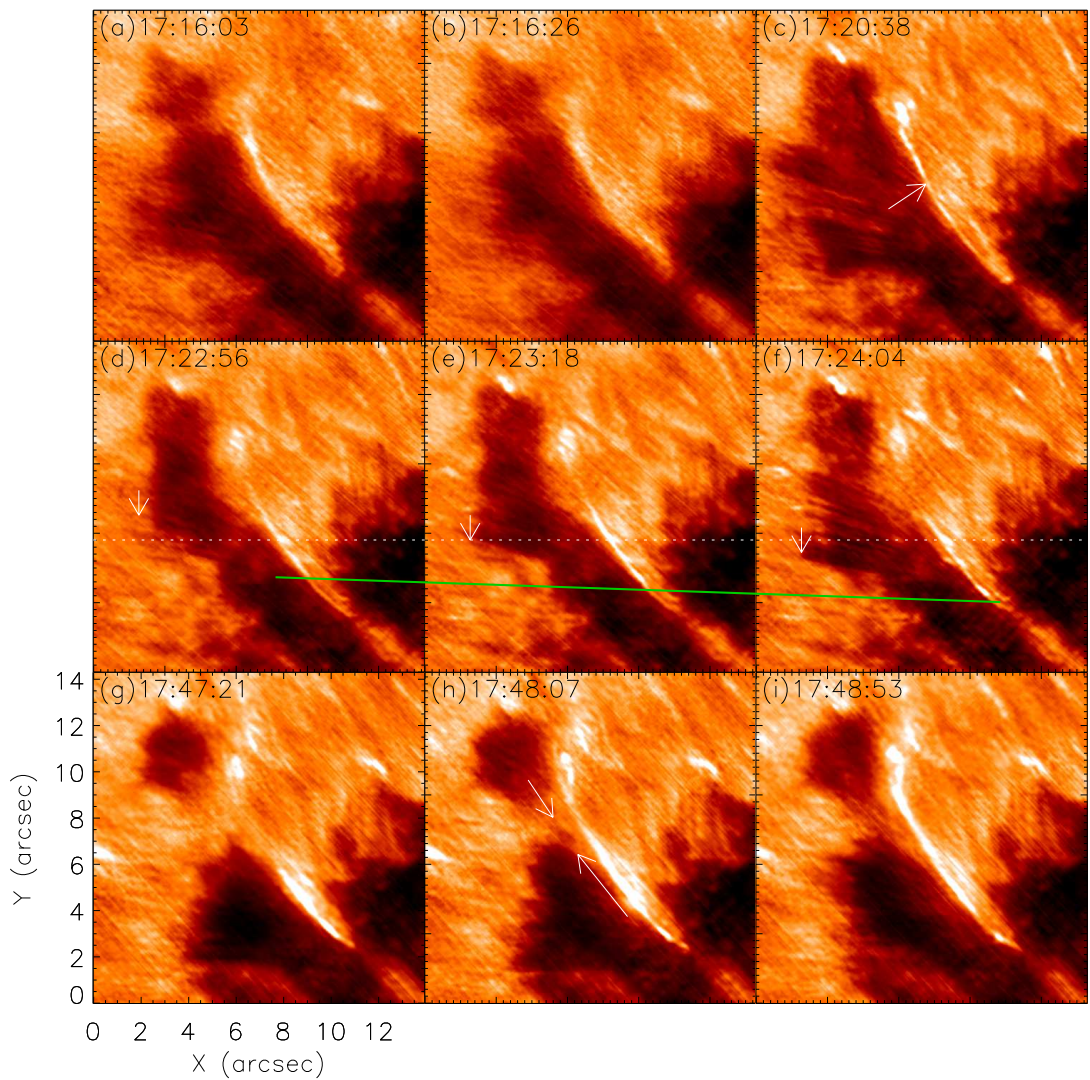} \caption{ Evolution of the H$\alpha$ off-band images at -0.8 {\AA} off the line center. The white arrow in panel (c) marks a ribbon-like brightening on LBs, associated with the fan shape jets. The green solid line illustrate the brightening motion on LBs along with the jets' motions perpendicular to their moving direction, which is possibly an indication of slipping reconnection. Two opposite arrows in (h) show the moving direction of the brightenings at the both ends of the LBs, which finally forms a complete ribbon. An associated animation (movie2.mov) is available online.}
\label{fig3}
\end{figure}

\begin{figure}
\epsscale{1.0} \plotone{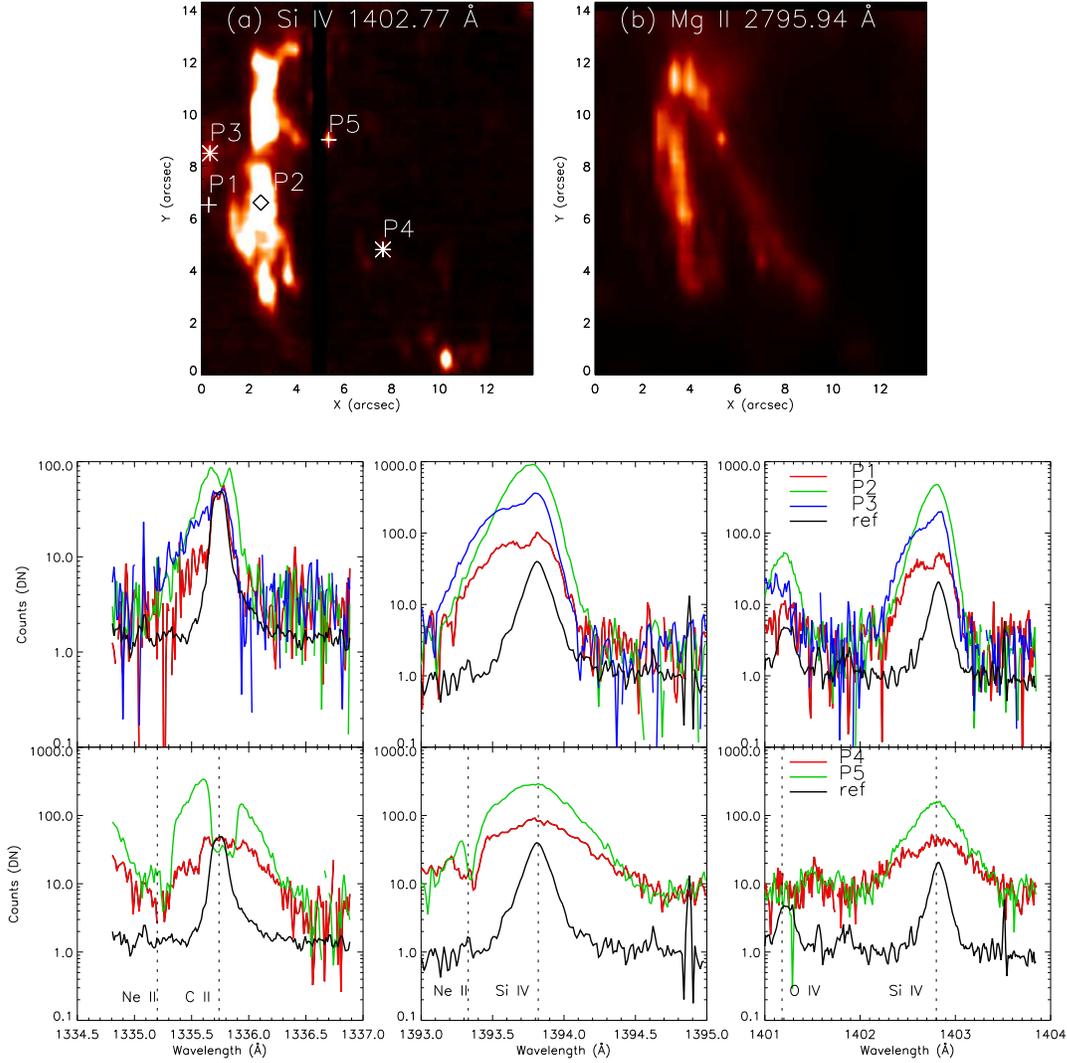} \caption{ Top row shows the slit scanning maps at the line centers of the Si {\sc iv} 1402.77 {\AA} and the Mg {\sc ii} K2V lines. From left to right in the lower two rows, they represent the spectra extracted from the three points at the bright fronts of jets (upper panel) and the two points at their footpoints (bottom panel) in the lines of C {\sc ii} 1335, Si {\sc iv} 1394 and 1403 {\AA}, respectively. The black curve is the reference spectrum from the average of a region of $12\arcsec\times2\arcsec$ in the slit scanning image as shown in the top row panels.  }
\label{fig4}
\end{figure}

\begin{figure}
\epsscale{1.0} \plotone{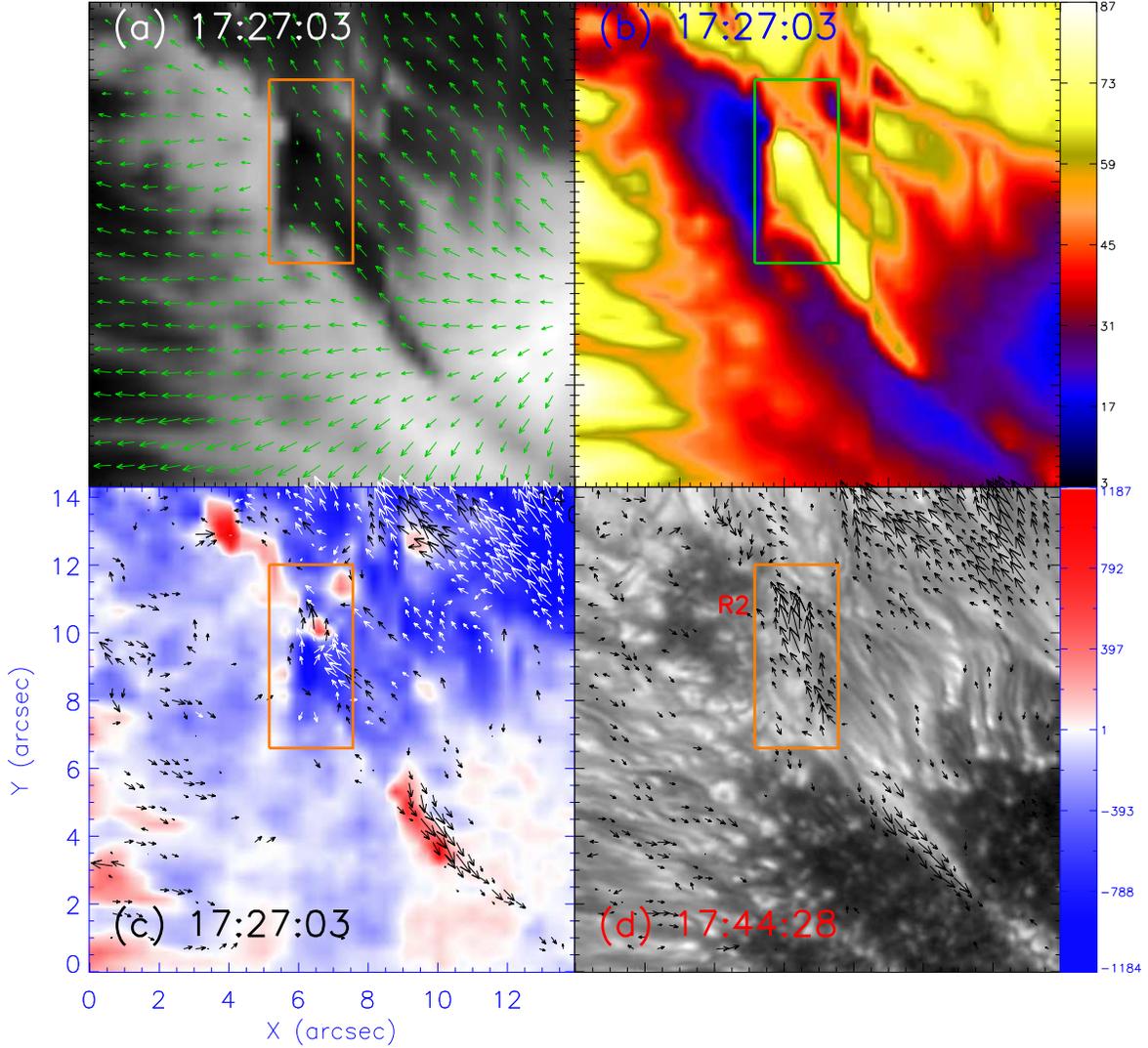} \caption{ (a): Longitudinal magnetogram in region R1 as shown in Fig. \ref{fig1}. The arrow represents the length and direction of the transverse magnetic fields. (b): Inclination angle map. (c): LOS velocity map from the SP data near 17:27 UT. (d): TiO images at 17:44 UT from GST. The arrows in panel c and d mark the horizontal velocity derived from series of TiO images. The longest arrow represents 1.8 km$\ s^{-1}$. R2 is a selected region to be used in the following figure. An associated animation (movie3.avi) is available online.}
\label{fig5}
\end{figure}

\begin{figure}
\epsscale{1.0} \plotone{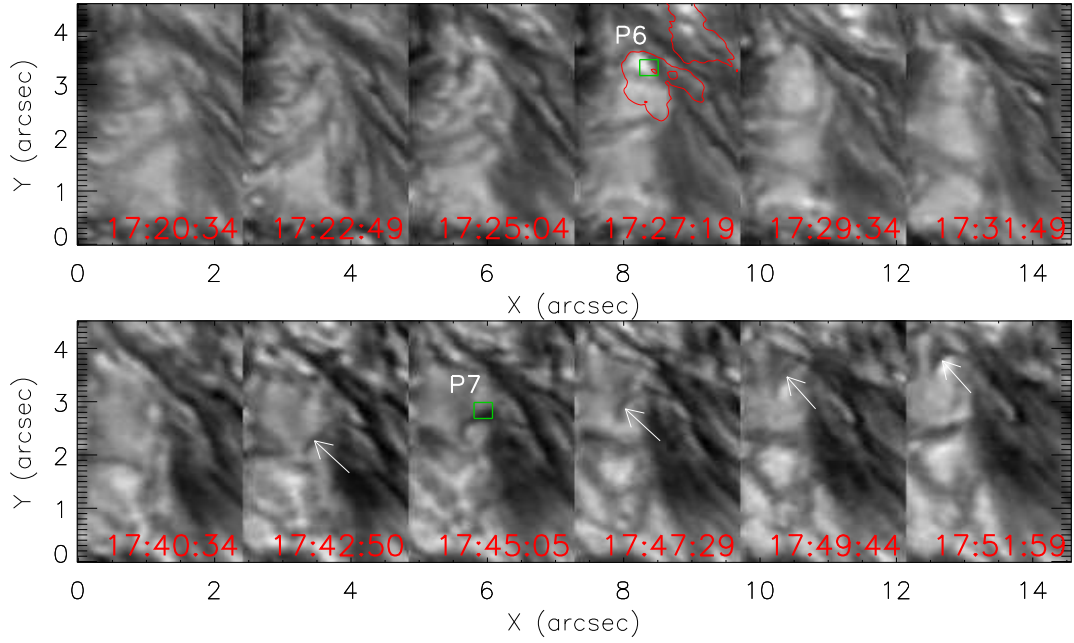} \caption{ Evolution of the TiO images in R2, the rectangle region shown in Figure \ref{fig5}. The red contours in the upper row mark the regions where we see brightenings in the Ha -0.8 {\AA} passband image. P6 and P7 are the locations where the abnormal Stokes V profiles are found. Those two locations have the same size as two pixels of Hinode/SP and their corresponding Stokes profiles are shown in Figure \ref{fig7}. White arrows in the bottom row mark the positions of a dark lane in fast motion.  }
\label{fig6}
\end{figure}

\begin{figure}
\epsscale{0.8} \plotone{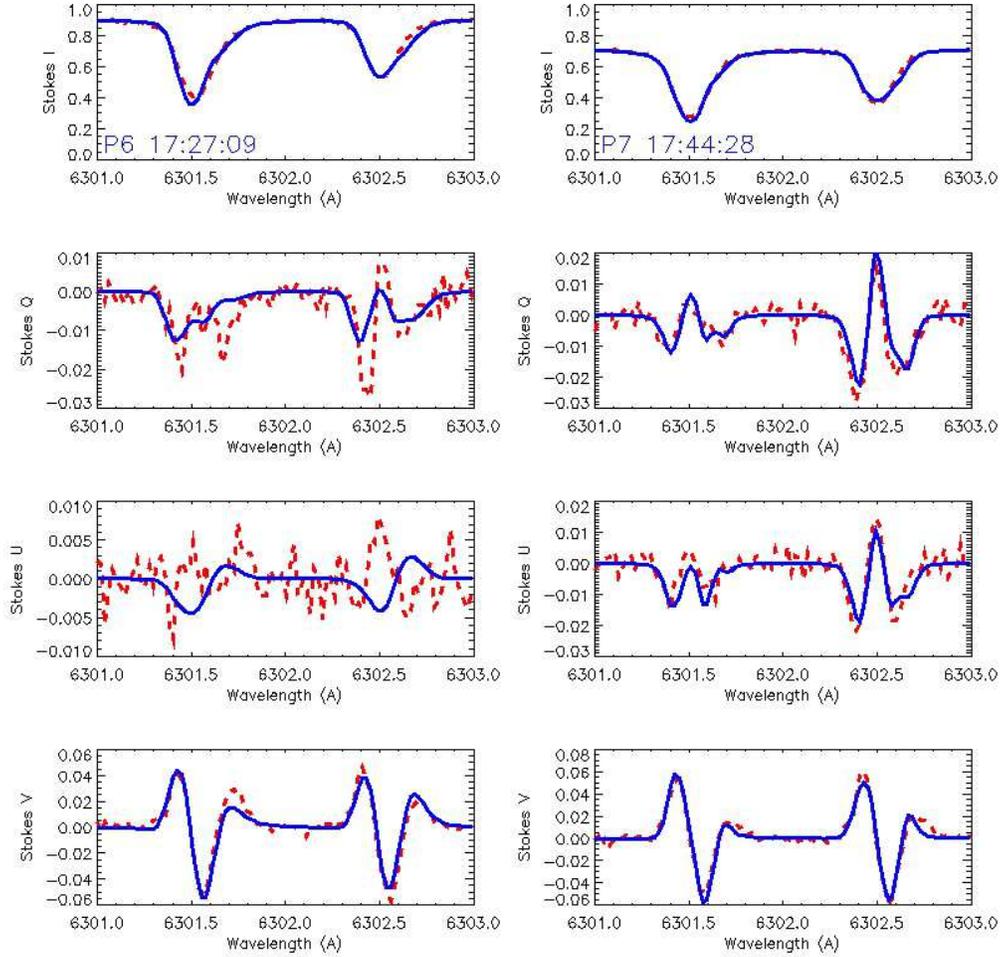} \caption{Stokes I, Q, U and V profiles in the two locations shown in Figure \ref{fig6}: P6 (left column) and P7 (right column). Red dashed line are the observational profiles from Hinode/SP while blue curves are the fitting results obtained using the NICOLE code.}
\label{fig7}
\end{figure}

\begin{figure}
\epsscale{0.8} \plotone{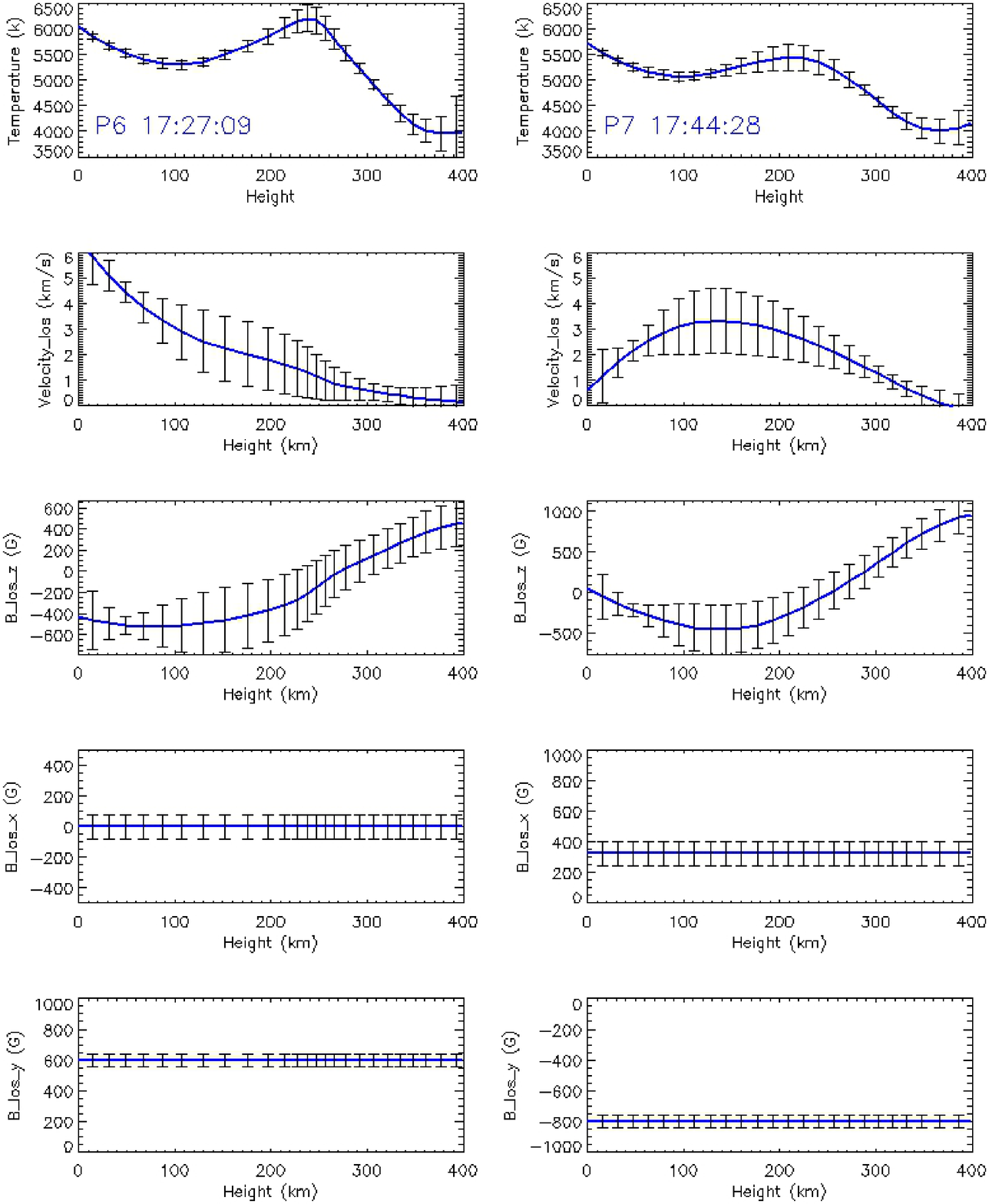} \caption{ Output model and corresponding estimated error obtained with the NICOLE code for the regions P6 (left column) and P7 (right column). From top to bottom, it is shown the stratification with height of temperature, LOS velocity, and the LOS of the three components of the magnetic field, namely, Bz, Bx and By.  }
\label{fig8}
\end{figure}

The length of the fan shape jets analysed here is about 4 Mm, in the range of the classification of the magnetic reconnection caused jets from \citet{Tian2018}. Numerical simulations are indeed powerful to understand the fan shape jets activities on LB. Unfortunately, the simulation from \citet{Toriumis2015} of an emerging LB does not contain the upper solar atmosphere so it is not able to investigate the dynamics of the jets and the physical mechanisms behind them. \citet{NS2016} and \citet{NS2017} carried out a 2.5D numerical experiments in which a surge results from magnetic flux emergence and the simulation contains the photosphere, chromosphere, TR and corona. The results from their simulation can explain some of the observations in the paper. But the simulation is not carried out in terms of LBs. Also due to its 2.5D nature, it can not recover all of the observations in the paper, especially in the transverse motions of the jets and the ribbon like brightenings seen in Fig. \ref{fig3}. More works need to be done on the simulation of fan shape jets on LBs and our results can provide some useful information to constrain the physical parameters of the simulation.

\acknowledgments  We thank the BBSO's observing and engineering staff for support and observations. BBSO operation is supported by NJIT and US NSF AGS-1821294 grant. GST operation is partly supported by the Korea Astronomy and Space Science Institute and Seoul National University and by the strategic priority research program of Chinese Academy of Science (CAS) with Grant No. XDB09000000. We acknowledge the free data usage policy of the SDO. IRIS is a NASA small explorer mission developed and operated by LMSAL with mission operations executed at NASA Ames Research center and major contributions to downlink communications funded by ESA and the Norwegian Space Centre. We also thank the Community Spectropolarimetric Analysis Center for providing the HINODE/SP data. This research work is supported by the Grants: 11873062, 11427901, 11427803,11773072, 11773038, 11773040, 11703042,U1731241, 11573012, 11673038, XDA15010800, XDB09040200 and US NSF AGS-1821294. The work is also supported by the Young Researcher Grant of National Astronomical Observatories, Chinese Academy of Sciences.

\end{document}